# Off-Lattice Self-Learning Kinetic Monte Carlo: Application to 2D Cluster Diffusion on the fcc(111) Surface

Oleg Trushin[1], Handan Yildirim[2], Abdelkader Kara[2] and Talat S. Rahman[2]
[1]Institute of Physics and Technology of RAS, Yaroslavl branch, Yaroslavl 150007, Russia
[2] Department of Physics, University of Central Florida, Orlando, FL, 32816-2385, USA

**Abstract**

We report developments of the kinetic Monte Carlo (KMC) method with improved accuracy and increased versatility for the description of atomic diffusivity on metal surfaces. The on-lattice constraint built into our recently proposed Self-Learning KMC (SLKMC) [1] is released, leaving atoms free to occupy "Off-Lattice" positions to accommodate several processes responsible for small-cluster diffusion, periphery atom motion and hetero-epitaxial growth. The technique combines the ideas embedded in the SLKMC method with a new pattern recognition scheme fitted to an Off-Lattice model in which relative atomic positions is used to characterize and store configurations. Application of a combination of the "drag" and the Repulsive Bias Potential (RBP) methods for saddle points searches, allows the treatment of concerted cluster, and multiple and single atom motions on equal footing. This tandem approach has helped reveal several new atomic mechanisms which contribute to cluster migration. We present applications of this Off-Lattice SLKMC to the diffusion of 2D islands of Cu (containing 2 to 30 atoms) on Cu and Ag(111), using interatomic potential from the Embedded Atom Method. For the hetero system Cu/Ag(111), this technique has uncovered mechanisms involving *concerted* motions such as shear, breathing and commensurate-incommensurate occupancies. Although the technique introduces complexities in storage and retrieval, it does not introduce noticeable extra computational cost.

# I. Introduction

The trends in modern studies of material designs follow a multi-scale and multi-disciplinary approach in which knowledge from various perspectives and at different length and time scales is brought together in as seamless a manner as possible. In these multi-scale approaches, the goal is to use filtered information obtained at the microscopic level to predict behavior at the macroscopic one following several well designed steps. Since materials evolve in time and space, the crucial first step is an accurate determination of the potential energy landscape and dynamics of the system at the atomistic scale. For well-designed relatively small systems this may often be done using *ab initio* electronic structure calculations [2] which being computationally intensive usually limit considerations to the equilibrium configuration at 0 K and no temperature or time evolution of the system is offered. *Ab initio* molecular dynamics simulations [3], on the other hand, can track temporal and thermal evolution of systems, albeit for limited length and time scales. To carry out simulations of large scale systems (containing more than few hundred atoms) that undergo evolution in temperature and time (beyond picoseconds), one usually resorts to the usage of many-body interatomic potentials. These model potentials allow one to simulate the behavior of materials along two different but often complementary paths. The first involves an explicit inclusion of the thermal motion of all particles in the system, via classical molecular dynamics (MD) simulations. These types of simulations are in general limited to time scales of microseconds (at best so far) that are still orders of magnitude smaller than those of experiments. One of MD's advantages is the explicit inclusion of system vibrational and thermal dynamics as controlled by the chosen interatomic potential and rules of classical mechanics. Together they allow the system to evolve "freely" (as a micro-canonical system) for timescales (few hundred nanosecond to microseconds) whose limit is imposed by available computational resources. Since typical experiments of interest (epitaxial growth, for example) report important morphological changes in time scale of minutes and hours, MD simulations may not be able to capture critical rare events and thereby fail to provide accurate time evolution of the system. The second path is to use Kinetic Monte Carlo (KMC) simulations, in which the thermal motion of the system is included only implicitly and in an approximate way. These simulations can cover time scales of minutes, hours or even days. In KMC, rates of "allowed" processes, through which the system evolves, are provided as an input [4-6]. The challenge is to determine as accurately as possible the parameters associated with these processes; and as completely as possible the list of allowed processes for the system, both being challenging, if not improbable, tasks. Accuracy in the determination of the energetics alone (neglecting the dynamical part) requires significant computational resources, and completeness is problematic, since evolution of the system according to a set of "allowed" processes may foster hitherto unforeseen rate controlling events. Several efforts have been devoted to overcome both handicaps when studying spatio-temporal evolutions of epitaxial and thin film growth [7-9]. One of the promising set of studies was carried out by Voter *et al* [10, 11] who focused on enhancing the time scales achievable in MD simulations through three different strategies: parallel-replica, temperature-accelerated dynamics and hyper-dynamics. In the same vain, Fichthorn *et al* [12] proposed a method which consisted in boosting the interaction energy in order to accelerate thermally activated processes. One may note that these two approaches are based on the boosting of a deterministic time evolution of the system. Another approach, which is probabilistic, has focused on the completeness issue of KMC by allowing the system to evolve according to a sequence of transitions with no "a-prioriness". The work by Jonsson *et al* [13], for example, carries out an extensive search of "possible" mechanisms and has successfully revealed several novel ones. The technique however, has some limitations. First, the "search" has to be limited to a pre-fixed number of processes and second, the knowledge gathered in any given search is wasted and hence, redundant calculations are performed throughout the simulations. We have recently introduced a method [1] that addresses both issues, namely the completeness and the redundancy. In this self-learning KMC (SLKMC) technique, a pattern-recognition scheme allows efficient storage and subsequent retrieval of information from a database of diffusion processes, their paths and their activation energy barriers. The method is based on the assumption that all atoms in the system occupy high-symmetry sites commensurate with the substrate (on-lattice sites). This was a good assumption for the study of the diffusion of relatively large 2D adatom islands on Cu(111) and Ag(111) surfaces [14-15], but ran into difficulties when island atoms moved away from lattice sites. For example, 6-atom and 10-atom islands of Cu on Cu(111) present "meta-stable" states which are characterized by the presence of irregular atomic positions (Fig.1). This is even more dramatic for hetero-systems in which atoms for islands of all sizes may occupy "Off-Lattice" sites. An accurate description of diffusive processes in such systems thus requires a more advanced pattern-recognition scheme allowing atoms to occupy any position on the surface. The approach described in this paper incorporates these new requirements and hence called Off-Lattice SLKMC. The idea is to replace the pattern recognition-scheme in SLKMC in which on-lattice occupancy is identified (using a binary scheme), by a new one in which "local" shapes are recognized and categorized according to the relative positions of the atoms of interest using real numbers.

The paper is organized as follows. Section II presents a short description of the model used in the SLKMC algorithm, the new pattern-recognition scheme, and accompanying methods for saddle point searches. In section III, we describe our finding for the diffusivity of a 6-atom Cu island on Cu(111) using this Off-Lattice SLKMC and compare our results with those previously published using SLKMC [14] and long-time molecular dynamics (MD) simulations [16]. In section III, we also present examples of novel and interesting mechanisms encountered during early stages of the simulation of the diffusion of Cu islands, with sizes ranging between 4 and 30 atoms, on Ag(111). Concluding remarks are presented in section IV.

**II. Model**

The system used in our simulations consists of a substrate and a cluster on top of it (Figure 2). The substrate consists of 5 atomic layers with 240-500 atoms per layer. Periodic boundary conditions are applied in the plane parallel to the surface of the substrate to mimic an infinite system. The two bottom layers of the substrate are fixed to prevent the sample from moving as a whole (when using molecular static simulations for saddle point searches). During KMC simulations, the substrate is an active participant in the evolution of the system. The interatomic potentials are modeled using the Embedded Atom Method [17], which has been shown to be reliable for calculations of self-diffusion on Cu and Ag surfaces [18].

**II.1. The SLKMC Algorithm**

The main idea incorporated in the SLKMC method is the usage of a pattern-recognition scheme to determine whether the energetics of the possible processes have been previously determined and stored during the course of the simulation. If all possible processes are stored, no further action is taken and the KMC simulation continues its course as standard KMC. But whenever a new configuration appears with new possible diffusion processes, a saddle point search is undertaken to complete the list of all possible processes for the configuration at hand. Once this procedure is finished and new processes along with their energetics are stored, the KMC simulation is resumed. We use the standard KMC algorithm to model sequences of diffusive events for a given system. We start from an empty database and search for the transition paths for each unknown local configuration.

We use several numerical methods to activate possible diffusive processes for a given cluster shape. A simple one, which we have named "**drag**" for obvious reasons, is both efficient and easy to implement. In this method we move each border atom in the direction of the nearest vacant site by small steps. During the process the system is relaxed in all directions except the one in which the atom is "dragged", thus preventing it from moving back to the initial state. Single and multi-atom processes may result from this method. The other method that we have found useful is based on **Repulsive Bias Potential (RBP)** [19]. In this method we start with small random displacements for all cluster atoms and then use a spherically symmetrical Repulsive Bias Potential to move the system from its initial local minimum to a neighboring one. The system is then allowed to fully relax in the modified energy surface.

All total energy minimization procedures are performed using a MD cooling (cooling to 0 K) method which has been described elsewhere [19]. The initial and final states for each process and the accompanying energy barrier are recorded in the database for further use. In order to be able to retrieve information from the database we use a unique characterization of the local configuration, which we describe below.

### II.2. Pattern-Recognition Routine

The central point of SLKMC algorithm is the pattern-recognition scheme. In order to avoid redundancy, the system "learns" from its "past" the processes and their energetic associated with previously encountered configurations. It is hence necessary to be able to distinguish different shapes of the system as they appear and make decisions using information already stored. The off-lattice model uses a set of relative positions of the atoms with respect to a chosen "reference point". We have chosen the reference point in two ways: 1) the closest fcc site to the "leading" atom in the system (note that any atom in the system can be selected to be the leading atom and in practice we choose the atom with label "1"); 2) the fcc site closest to the center of mass of the system. Every atom in the system then has uniquely defined coordinates with respect to this reference point as shown in Fig.3 (for case 1).

### II.3. Symmetry Operations

To avoid redundancy in repeating searches for symmetrically equivalent paths, we perform symmetry operations on every new process and store the results. For fcc(111) there are 5 possible symmetry operations: $120^o$ and $240^o$ rotations of the system, mirror reflection in the vertical plane and two combinations of the above-mentioned rotations with mirror reflections. Once for a novel configuration the search for paths is completed using the activation procedures discussed above and the information recorded, we create 5 symmetrical configurations using the above-mentioned symmetry operations.

## II.4. Detailed Balance Conditions

To satisfy detailed balance conditions we need to use the same transition path for the forward (from initial to final configurations) and backward (from final to initial configurations) processes. Doing so ensures that all configurations accumulated in the database are connected by unique transition paths and that starting from any arbitrary configuration and performing any number of processes to return to the initial configuration, using any possible path, would result in a zero energy difference between starting and ending energies. In practice, we proceed as follows: each search for a transition carries information about the initial, saddle and final configurations. The energies are then used to store the activation energies for both forward and backward processes. Very often we found that at a given time of the simulation no atoms in the systems were candidates for the stored backward processes. This is not a problem. Instead the inclusion of the backward mechanisms in the probabilistic procedure with zero probability ensures completeness and maintenance of detailed balance.

## II.5. Repulsive Bias Potential for Saddle Point Search

In this work we have applied the Repulsive Bias Potential (RBP) method [19] for searching transition paths for new configurations that appear in the KMC trajectory for the diffusing entity. The novelty in the current implementation is that the method is used not only for determining possible final states but also for saddle point location. In RBP procedure there are two parameters controlling the shape of the applied repulsive potential: 1) the amplitude of repulsion and 2) the rate of exponential decay [19]. The last parameter determines the effective region of the potential. Thus by changing this parameter we can change the region of the energy surface affected by the bias potential. To locate the saddle point we start the RBP procedure from a very narrow repulsive region and spread the repulsion gradually by changing the corresponding parameter in small steps until we reach the saddle point. This procedure is illustrated in Fig.4. The moment of saddle point crossing is indicated by changing the sign of the slope of the energy surface. At each step during the procedure we calculate the difference between the energy of the current configuration and of the initial state. During the climb to the saddle point, this quantity increases and after crossing saddle point it starts to decline, and the saddle point naturally corresponds to the configuration with maximum of this energy difference. After arrival at the final state we estimate the energy barrier associated with this transition path as a maximal value on the energy profile. Since there may be many possible transition paths starting from a given initial state, the RBP method selects only one of them using one set of small initial random displacement and parameters of the repulsive potential. To determine as many transition paths as possible, several searches using different sets of random displacements are performed.

## II.6. Database Format

In the original SLKMC, in the 3-ring format, three integers were enough to completely determine a given configuration. In the present database, on the other hand, a given configuration is determined by the relative positions (x,y,z) of all atoms, making the database much more storage demanding than that of SLKMC. Every configuration in the database is tagged as contributing either to "forward" or "backward" motion. Forward configurations are actually found during the course of the simulation and hence are "complete" in the sense that all possible (within the limitations of our saddle point search engines) mechanisms within this configuration have been determined. The backward configurations are "incomplete" in the sense that only mechanisms that take this final configuration to the initial configuration associated with it are stored. Each transition is characterized by the corresponding energy barrier and the displacement field that will take the system beyond the initial configuration by performing a given process. Thus diffusive motion of the cluster is achieved by adding a displacement field to the current coordinates of the adatoms.

## III. Results

## III.1 Diffusion of a 6-atom Island: Cu/Cu(111)

As we have mentioned earlier, the need for an Off-Lattice SLKMC came from observations made during the simulation of 6-atom cluster diffusion, where we found that for some shapes the atoms did not occupy on-lattice positions. Hence, we present here some details about the simulation of 6 copper atom diffusion on Cu(111) at 300K, using the off-lattice method described above. Before we show results of the diffusion, let us examine a couple of mechanisms resulting from the first 100 KMC steps. Indeed, the RBP revealed multi-atom mechanisms in which the atomic occupancy can be either "on" or off-lattice. In Fig.5a and b, we show two types of mechanisms. In the first the whole cluster diffuses from "fcc" to "hcp" occupancy (Fig.5a), while in the second (in Fig.5b), there is a "partial" rotation of the cluster (as shown by an arrow), in which two atoms (#1 and #2) do not move and stay "on-lattice," while the rest of the atoms (#3-#6) move (each with different displacement) resulting in "off-lattice" occupancy. This feature illustrates the new capabilities of the Off-Lattice SLKMC.

Note that these mechanisms were revealed automatically by the RBP, whereas in the case of SLKMC, collective motion mechanisms had to be added by hand [14]. A long KMC run at 300K involving 1 million steps was performed to study the diffusion of 6-atom clusters of Cu on Cu(111). In Fig. 6a, we show the trace of the center of mass (CM) of the island during the simulation which indicate that the island diffusion is of a random-walk type. This is also in agreement with the linear behavior of the time evolution of the CM mean-square-displacement, as shown in Fig. 6b. Using the proportionality between the slope in Fig. 6b and the diffusion coefficient, we extract a value for the latter of $7.8 \times 10^7$ $Å^2/s$, which is of the same of order of magnitude ($3 \times 10^7$ $Å^2/s$) as that found by Karim *et al* using SLKMC [14]. We also find that both of these KMC results are in good agreement with recently published values of diffusion of small Cu islands on Cu(111), using MD simulations, where it was found that diffusion was following an Arrhenius behavior with an extrapolated value at 300K for a 6Cu-atom island diffusion coefficient of about $4. \times 10^7$ $Å^2/s$ [16]. This result is remarkable given the fact that MD results were obtained using a modified EAM potential in addition to a different method for simulations: MD includes explicitly the vibrational dynamics of the system, whereas in KMC, the substrate is taken to be rigid and all vibrational dynamical information is implicitly "buried" in the prefactor (often taken to be $3 \times 10^{-3} cm^2/s$ for *all* mechanisms).

**III.2. Novel mechanisms in Cu/Ag(111): Diffusion of 4-30 atoms 2D Islands**

The examination of cluster diffusion for hetero-systems is rich, interesting and challenging. It offers both opportunities for uncovering interesting physics and challenges in their evaluation. Indeed, because of the lattice mismatch and different strengths in the bondings between like and unlike atoms in a variety of environments, the energy landscape on which the system evolves presents interesting features. Depending on the number of "like" neighbors vs. that of "un-like" neighbors, atoms in the cluster have a variety of occupancy. Hence, the off-lattice SLKMC is the only way to simulate the evolution of hetero-systems. For example, for the case of a Cu/Ag system, we note that Cu has a lattice constant of 3.615Å with a cohesive energy of 3.6eV, while Ag has a lattice constant of 4.09 Å and a cohesive energy of 3.0eV. The disparity between these two quantities is enough to yield the appearance of novel mechanisms for the diffusion of small 2D islands of Cu on Ag(111), as we shall see.

For cluster size ranging between 4 and 30 Cu atoms, we have analyzed the processes revealed by the saddle point search engines, incorporated in our Off-Lattice SLKMC, during the first 100 KMC steps at a temperature of 300K. The focus of this paper is to demonstrate the feasibility of this new method and explore its features and some of the results. Hence, we will show here only selected novel processes in diffusion of Cu small clusters on Ag(111).

We find that a 6Cu-atom cluster can glide as a whole across Ag(111), as we saw it do across Cu(111). But it can also diffuse in other ways, sometimes in combination with concerted motion of the whole, sometimes without. In Fig. 7a we show one of the ways in which shearing of the cluster can occur. The 6Cu-atom cluster can here be analyzed as a tetramer made of atoms #1-4 (on the left), to which a dimer is attached (atoms #5 and 6). The dimer shears along the side by which it attaches to the tetramer. Meanwhile, the tetramer itself slides from fcc to hcp sites with respect to the substrate. In Fig. 7b we show how a single peripheral atom (here #6) diffuses along a (100)-micro-facetted (A-type) step consisting of the Cu atoms labeled #1, #3 and #5. These two mechanisms were made possible by the relatively high frequency with which the cluster changes shape during the 6Cu-atom diffusion on Ag(111), in comparison with the relative rarity of such shape change in the case of 6Cu-atom diffusion on Cu(111).

For an 8Cu-atom cluster on Ag(111), we observed – in addition to the usual gliding of the whole cluster in which Cu-Cu distances remain fixed during the diffusion from fcc to hcp occupancy (and vice versa) – various shear mechanisms as well. The 8Cu-atom cluster (Fig. 8) behaves as a pair of tetramers – one consisting of atoms #1-4 and the other of atoms #5-8, the first of which diffuses along the side of the second, in the direction indicated by the arrow.

During the first 100 KMC steps of the diffusion of a13Cu-atom cluster on Ag(111), we often observed the concerted gliding of the whole cluster from fcc to hcp to fcc and so on. Surprisingly, though, the simulation turned up a pattern in which a pair of atoms at the edge of the cluster remained in place while the much larger remainder of the cluster underwent a "commensurate-incommensurate" transition illustrated in Fig. 9a. Although this mechanism exhibits very limited diffusivity, it can trigger shape change that leads in turn to the gliding of the whole cluster. In addition, we observed several diffusion processes in which a triangular subunit undergoes shearing motion. We present one example of such shearing in Fig. 9b, where the shearing triangle consists of atoms #11, #12 and #13.

The 22Cu-atom cluster exhibits a yet different behavior, which we illustrate here in a series of three snapshots that indicate a pair of consecutive processes that, taken together, might be characterized as constituting a "breathing" mode. We begin, in Fig. 10a, with an open-structured cluster with fully commensurate occupancy (all atoms occupying fcc sites). In Fig. 10b the left side of the cluster (atoms #1-#3) has moved towards the rest of the cluster, giving the whole a more compact shape, with the right part of the cluster occupying fcc sites while the left occupies hcp sites. Note that the A-type step on the left of the cluster in Fig. 10a (formed by atoms #1, #2, #3) turns into a B-type step in Fig. 10b. In the diffusion taking place between Figs. 10b and 10c, the atoms on the right side of the cluster move from fcc to hcp sites. Also note that the B-type step on the right of the cluster in Fig. 10b (formed by atoms #20,#21,#22) turns into an A-type in Fig.10c. The whole cluster now occupies hcp sites and assumes an open structure again. This is an fcc-to-hcp motion in two phases instead of a single concerted glide characteristic of smaller clusters. In Fig. 11, we show a striking mechanism of shear involving the motion of the middle part of the cluster while the sides of the cluster do not move: the 22Cu-atom cluster can be analyzed as a set of six chains (labeled here from 1 to 6) with the left part occupying fcc sites and the right part hcp sites. The shear consists of a translation of the two middle chains (3 and 4 in the illustration) which is quite remarkable, since shear is usually thought of as motion of a peripheral portion of a cluster along the edge a central portion.

A 30Cu-atom cluster presents several breathing and shear mechanisms. One additional pattern in particular attracted our attention. Illustrated in Fig. 12, it involves the first stage of "popping-up" an atom (#6) from the rest of the cluster. The transformation of a partial-hexagon in the upper left part of the cluster (atoms #1-5, on the left) gets transformed into a pentagon (Fig.12 on the right), the central atom of which (#6) occupies a higher z (vertical) position. In fact, the Off-Lattice SLKMC with the new pattern recognition is able to incorporate inter-layer diffusion, provided that the saddle-point search engines associated with it are able to reveal these mechanisms.

**III.3. Checks of Calculated Energy Barriers**

As mentioned earlier, hetero-systems present a rich energy landscape, and our RBP method in its present form is incapable to adjust (automatically) the parameters for energy landscapes with diverse characteristics (for example, both steep valleys and shallow wells). As a consequence, energy barriers for some of the novel mechanisms may not be very accurately determined by the present RBP method. In order to take stock of this limitation, we systematically filtered *all* the energy barriers in the hetero-systems through a calculation of the same using the Nudged Elastic Band (NEB) method [13] which has come to be accepted as a reliable procedure for such calculations, albeit computationally intensive. As an illustration, we display in Table 1 the activation energy for some mechanisms of a 10-atom Cu cluster on Ag(111) as calculated using RBP and NEB methods.

**Table 1.** Activation energy of some chosen processes

| Process # | Barrier (eV) RBP | Barrier (eV) NEB |
|---|---|---|
| 1 | 0.395 | 0.355 |
| 2 | 0.166 | 0.158 |
| 3 | 0.004 | 0.005 |
| 4 | 0.174 | 0.167 |
| 5 | 0.148 | 0.167 |
| 6 | 0.568 | 0.883 |

**IV. Conclusions**

We have presented a novel development of Self-Learning KMC technique designed to improve accuracy of simulation of diffusive processes on metal fcc(111) surfaces. The novelty of the approach consists in using a Off-Lattice model in pattern recognition scheme. This enables us to make simulations more realistic by taking into account metastable states with irregular (off-lattice) positions of atoms in two-dimensional clusters. We tested the method by estimating the diffusivity of small two-dimensional clusters for two systems: Cu/Cu(111) and Cu/Ag(111). The results showed good agreement with those of MD simulations. This approach might be useful in studies of morphological evolution on metal surfaces and in further developments in KMC simulations of thin film growth.

**Acknowledgements**

This work was supported by NSF-ITR grant no. 0428826.

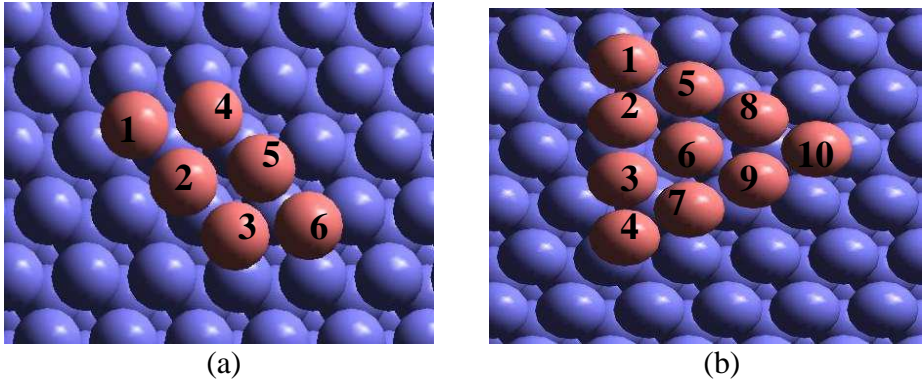

**Figure 1**. Metastable states for Cu/Cu(111) islands with irregular positions with respect to adatoms. **(a)** 6-atom island. **(b)** 10-atom island.

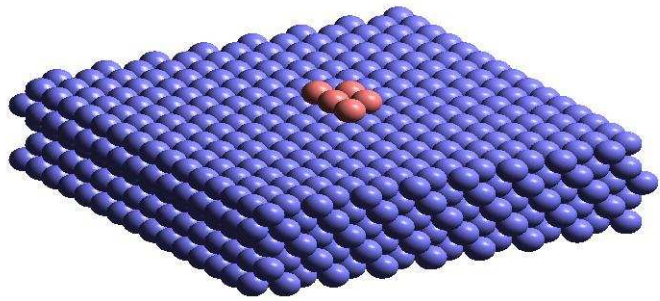

**Figure 2**. Example of model system

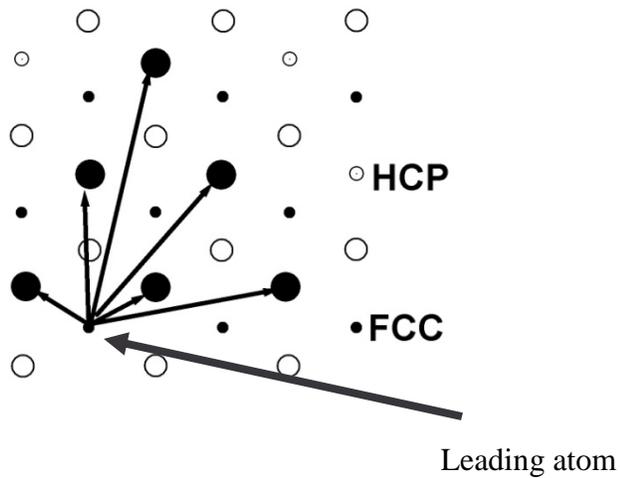

**Figure 3.** Off-Lattice Pattern-recognition scheme Positions of atoms are characterized by relative coordinates with respect to the fcc site nearest to the "leading atom."

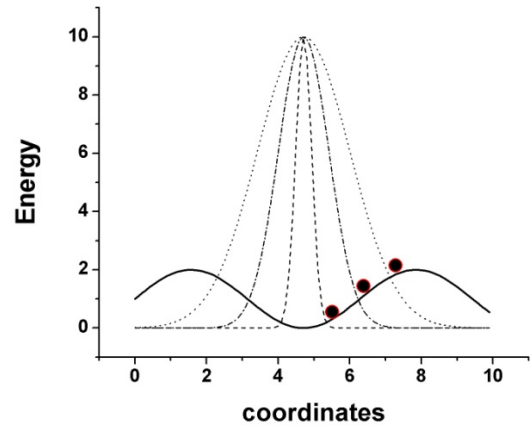

**Figure 4.** Illustration of Repulsive Bias Potential (RBP) used for saddle-point search.

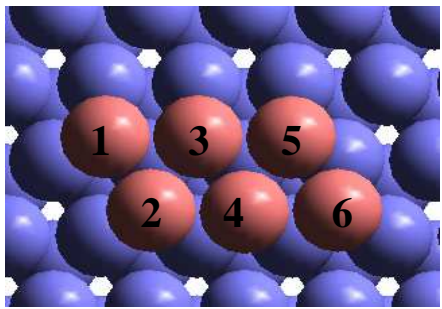 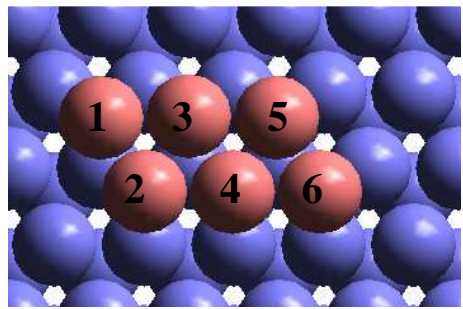

(a)

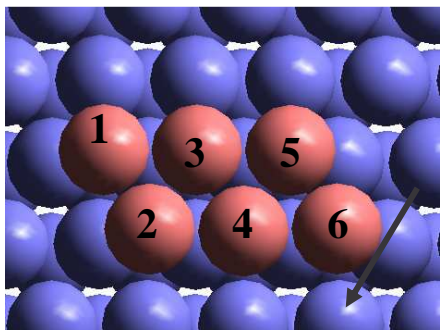 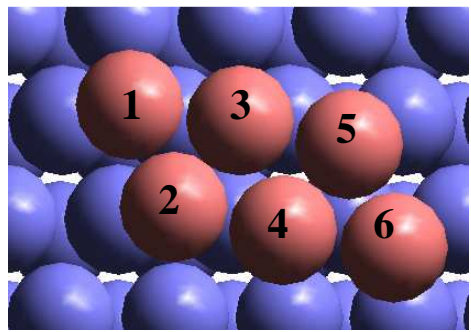

(b)

**Figure 5.** 6Cu-atom cluster on Cu(111). **(a)** Concerted motion of the cluster from fcc to hcp. **(b)** Partial rotation where only atoms #4, 5 and 6 moved from their initial positions, with different displacements.

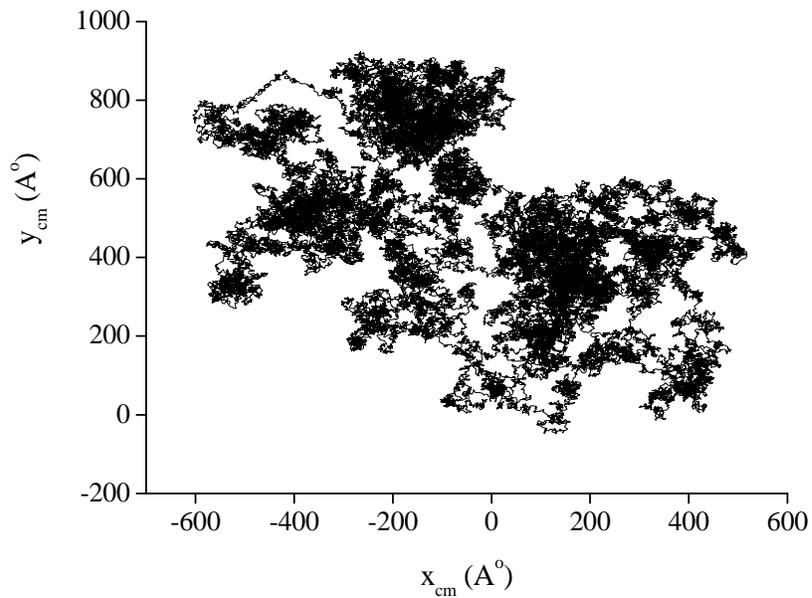

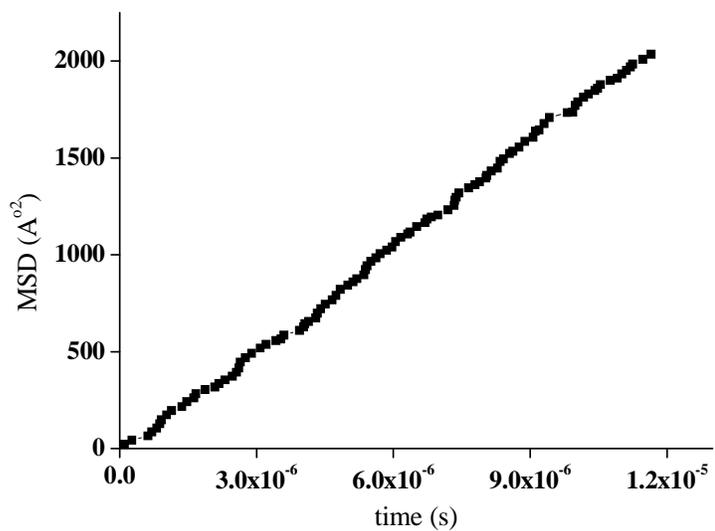

**Figure 6.** Simulation of 6Cu-atom cluster on Cu(111) at 300 K for 1 million steps. **(a)** Trace of center of mass of the cluster. **(b)** Mean square displacement (MSD) of the center of mass of the cluster.

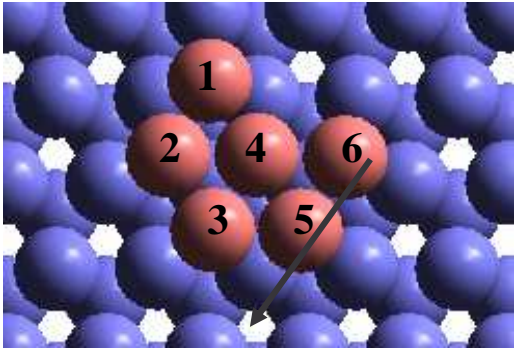
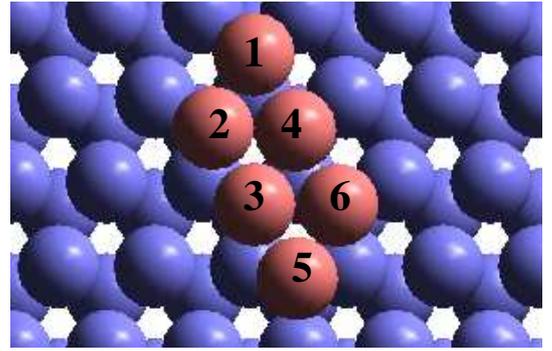

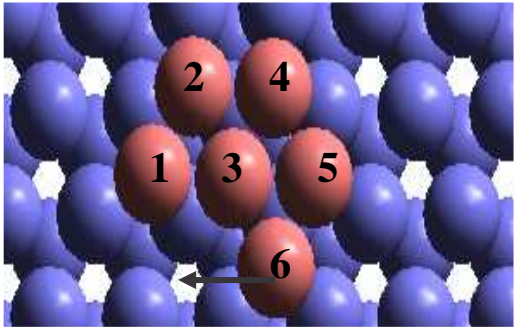
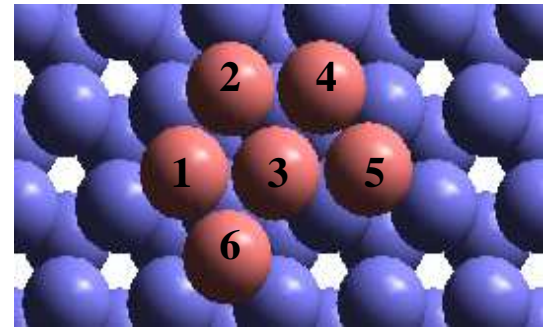

(a)

(b)

**Figure 7.** 6Cu-atom cluster on Ag(111). **(a)** Shearing diffusion of atoms #5 and 6, combined with concerted motion of the entire cluster from fcc to hcp. **(b)** Peripheral diffusion of atom #6.

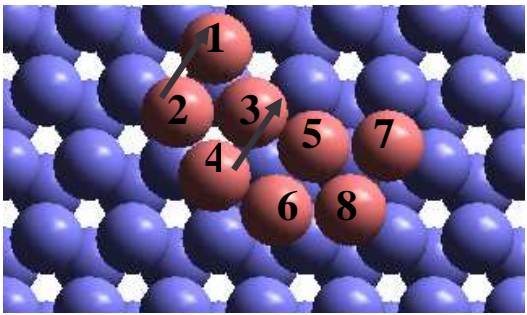
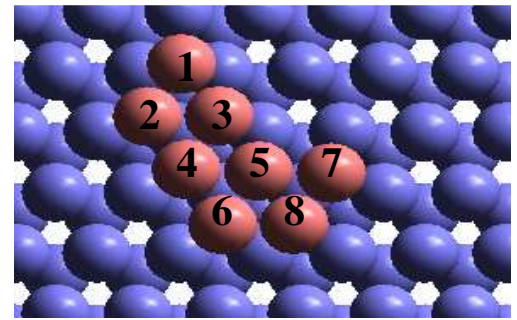

**Figure 8.** 8Cu-atom cluster on Ag(111): Shearing by the (#1-4) tetrameter.

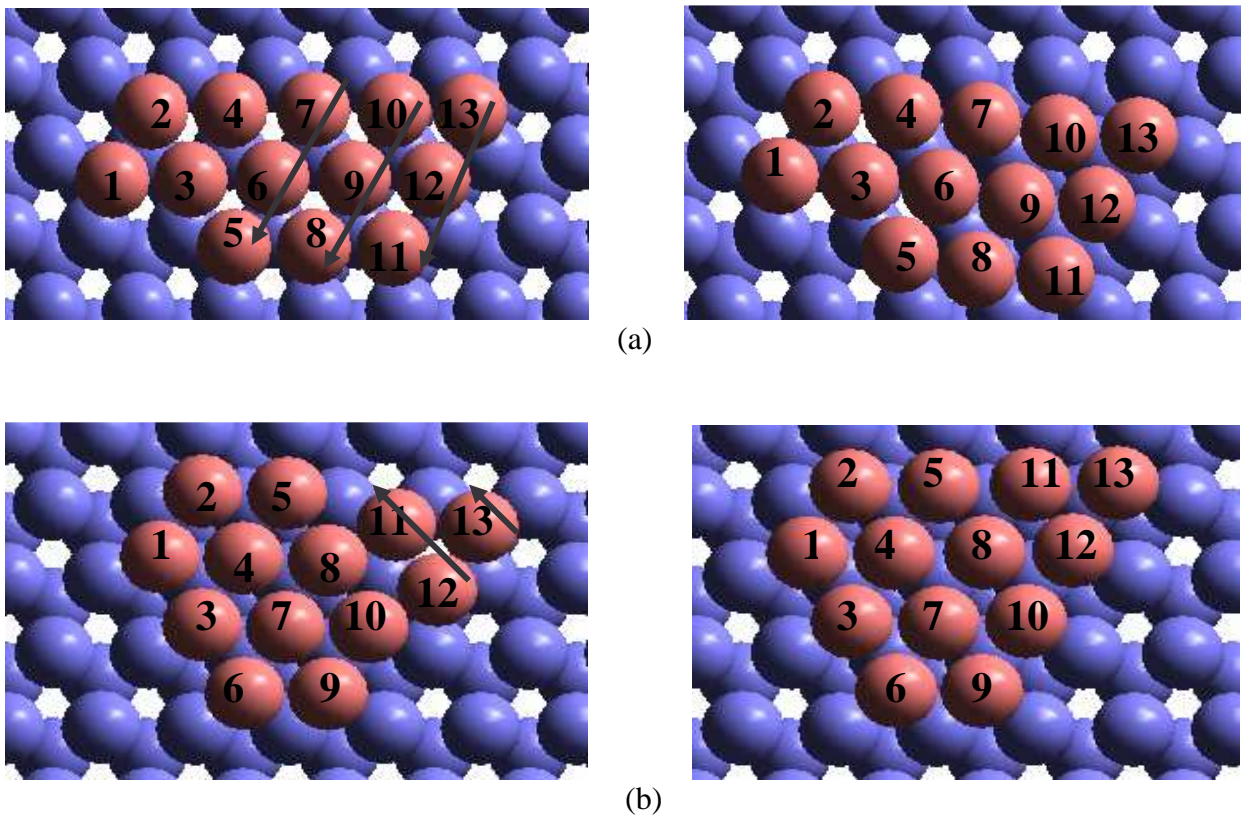

**Figure 9.** 13Cu-atom cluster on Ag(111). **(a)** Anchoring of two atoms #1 and 2 at the edge of the cluster, while the remainder shears. **(b)** Shearing by the triangle formed by atoms (#11-13).

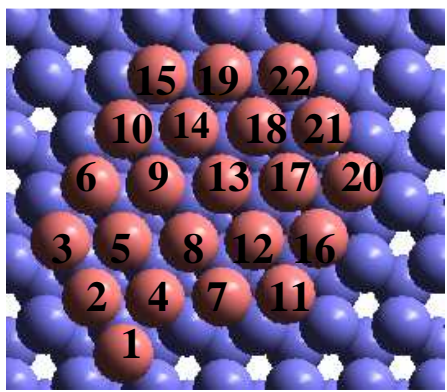

(a)

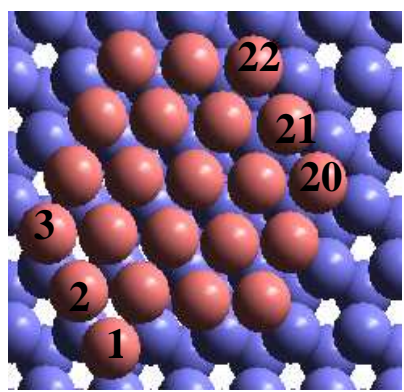

(b)

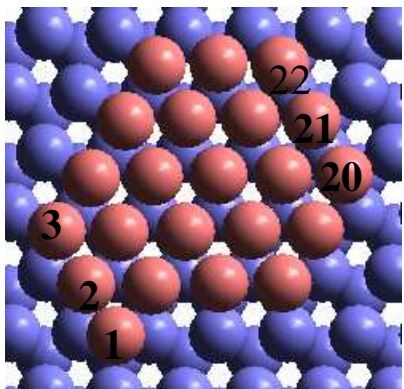

(c)

**Figure 10.** 22Cu-atom cluster on Ag(111): Breathing mode (see the text).

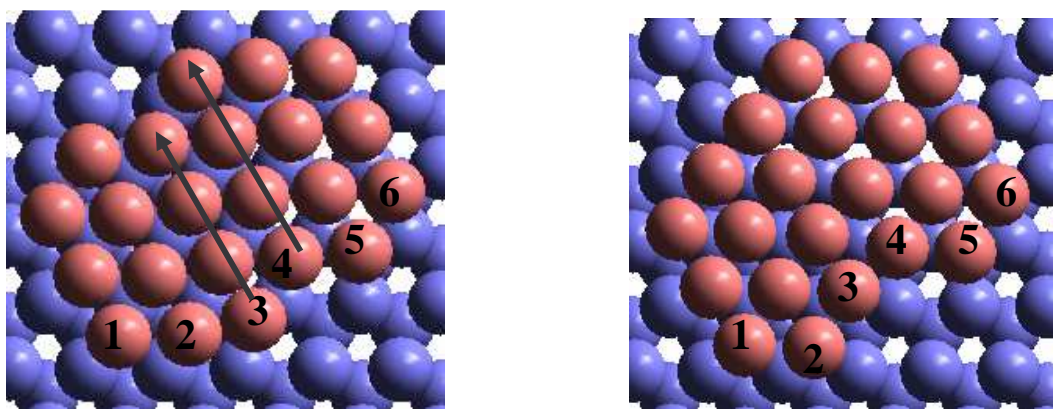

**Figure 11.** 22Cu-atom cluster on Ag(111): Shearing of the 3$^{rd}$ and 4$^{th}$ atom-chains.

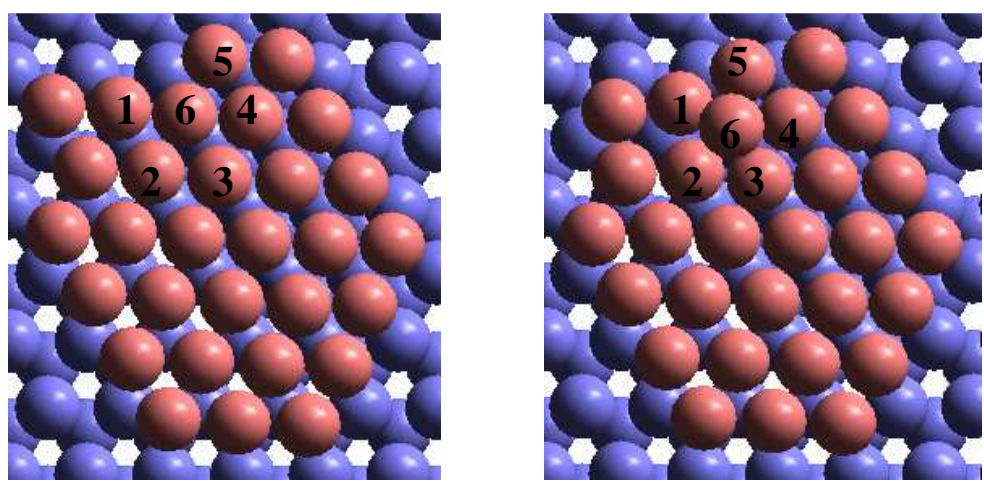

**Figure 12.** 30Cu-atom cluster on Ag(111): popping-up of an atom (#6).